\documentstyle[multicol,aps,epsf]{revtex} 
\begin{document}
\draft
 
\title{Fisher Waves in the Diffusion-Limited Coalescence Process
$A+A\rightleftharpoons A$}
\author{Daniel ben-Avraham\footnote
  {{\bf e-mail:} qd00@polaris.clarkson.edu}}

\address{Physics Department, and Clarkson Institute for Statistical
Physics (CISP), \\ Clarkson University, Potsdam, NY 13699-5820}
\maketitle

\begin{abstract} 
Fisher waves have
been studied recently in the specific case of diffusion-limited reversible
coalescence, $A+A\rightleftharpoons A$, on the line.   
An exact analysis of the particles concentration showed that
waves  propagate from a stable region to an unstable region at constant
speed, just  as in Fisher's ``mean-field" theory;
but also that the wave front fails to retain its initial shape and instead it
broadens with time.  Our present analysis encompasses the
full hierarchy of multiple-point density correlation functions, and thus it
provides a complete exact description of the same system.  We find that as
the wave propagates, the particles in the stable phase remain distributed
exactly as in their initial (equilibrium) state.  On the other hand, the
leading particle---the one at the edge of the wave---advances as a biased
random walk, rather than simply linearly with time.  Thus the shape of the
wave remains actually constant, but it is the ``noisy" propagation of the
wave's edge that causes its apparent broadening.  
\end{abstract}

\begin{multicols}{2}
Fisher waves~\cite{Fisher,Kolmogorov} are the best known paradigm of the
invasion of an unstable phase by a stable phase.  Fisher's theory describes
the kinetics of particles concentration $\rho(x_1,x_2,\dots,x_d,t)$ in a
reaction process  (taking place in $d$-dimensions) at the level of a
reaction-diffusion equation~\cite{Fisher,Kolmogorov}:
\begin{equation}
\label{Fisher.eq}
{\partial\rho\over\partial t}=D\Delta\rho+k_1\rho-k_2\rho^2\;.
\end{equation}
Here $D$ is the effective diffusion constant of the particles, and $k_1$ and
$k_2$ are rates of particle generation and particle death, respectively. 
Eq.~(\ref{Fisher.eq}) admits two stationary solutions:
\begin{equation}
\label{rho=0}
\rho=0\;,
\end{equation} 
and
\begin{equation}
\label{rho=k1/k2}
\rho=k_1/k_2\;,
\end{equation}
but only the latter is stable.
Suppose that the system is prepared at the initial state: $\rho=k_1/k_2$ for
$x_1\leq 0$, and $\rho=0$ for $x_1>0$.  Then the stable phase
($\rho=k_1/k_2$) invades the unstable phase ($\rho=0$) in the form of a wave,
$\rho(x_1,\dots,x_d,t)=f(x_1-ct)$, which travels unchanged at
constant speed
$c\geq c_{\rm min}=2\sqrt{k_1D}$.  The minimal speed is realized for
sufficiently sharp initial interfaces~\cite{F3} (such as in our case).  

Fisher's equation might be regarded as a ``mean-field" approximation to the
kinetics of diffusion-limited coalescence, $A+A\rightleftharpoons A$: The
reaction terms of Eq.~(\ref{Fisher.eq}) are in the form of the mass action
rate equation  appropriate for systems in local equilibrium, as might be
expected for reaction-limited kinetics.  Recently, however, 
diffusion-limited reversible coalescence was analyzed {\em exactly\/} in one
dimension~\cite{DBH}.  It was found that, just as in the Fisher theory, waves
propagate at a constant speed from a stable to an unstable phase, but that
the width of the wave front broadens with time as $w\sim\sqrt{t}$.  The
broadening of the wave front is of special interest since it represents the
effects of the internal noise in the system---the noise which Fisher's
equation fails to model.

In ref~\cite{DBH} exact expressions were derived for
the particle {\em concentration\/}, following the method of Inter-Particle
Distribution Functions (IPDF).  In this letter, we exploit the same method
to derive the {\em full distribution\/} of particles, as represented by the
infinite hierarchy of $n$-point density correlation functions. (The particle 
concentration corresponds to the special case of $n=1$.) We find that as
the wave propagates, the particles in the stable phase remain distributed
exactly as in their initial (equilibrium) state.  On the other hand, the
leading particle---the one at the edge of the wave---advances as a biased
random walk, rather than simply linearly with time.  In this view, the shape
of the wave remains actually constant in any particular realization of the
process.  The apparent broadening is the result of the
fluctuations in the location of the wave's edge among different realizations
(Fig.~1).

The coalescence model~\cite{DBH,Coalescence,Doering,AAreviews} is defined on
a one-dimensional lattice of lattice spacing
$a$.  Each site is in one of two states: occupied by a particle $A$, or empty.
Particles hop randomly into nearest neighbor sites, at rate
$D/a^2$.  A particle may give birth to an additional particle, into a
nearest neighbor site, at rate $v/a$ (on either side of the
particle)~\cite{remark}.  If hopping or birth occurs into a site which is
already occupied, the target  site remains occupied.  The last rule means
that coalescence,
$A+A\to A$, takes place {\it immediately\/} upon encounter of any two
particles.  Thus, together with hopping and birth, the system models the
diffusion-limited reaction process
$A+A\rightleftharpoons A$.  The system's dynamical rules are illustrated in
Fig.~2.

\begin{figure}
\centerline{\epsfxsize=5cm \epsfbox{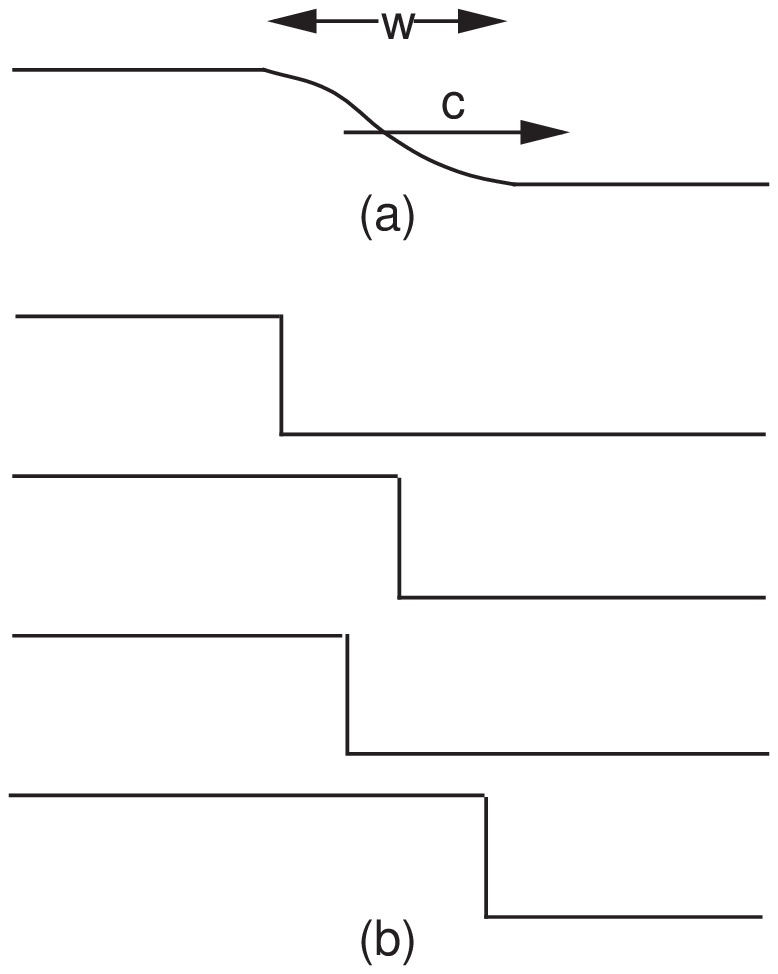}}
\noindent
{\small {\bf Figure~1}. Fisher waves: (a) The ensemble average yields a
wave traveling at speed $c=v$, with a front that  broadens with time as
$w\sim\sqrt{Dt}$.  (b) Individual waves actually remain
sharp throughout the motion, but their fronts travel in an erratic fashion,
leading to the ensemble average in (a).  Fisher's classical theory predicts
a wave as in (a), but its front does not broaden with time.}
\end{figure} 
\begin{figure}
\centerline{\epsfxsize=6cm \epsfbox{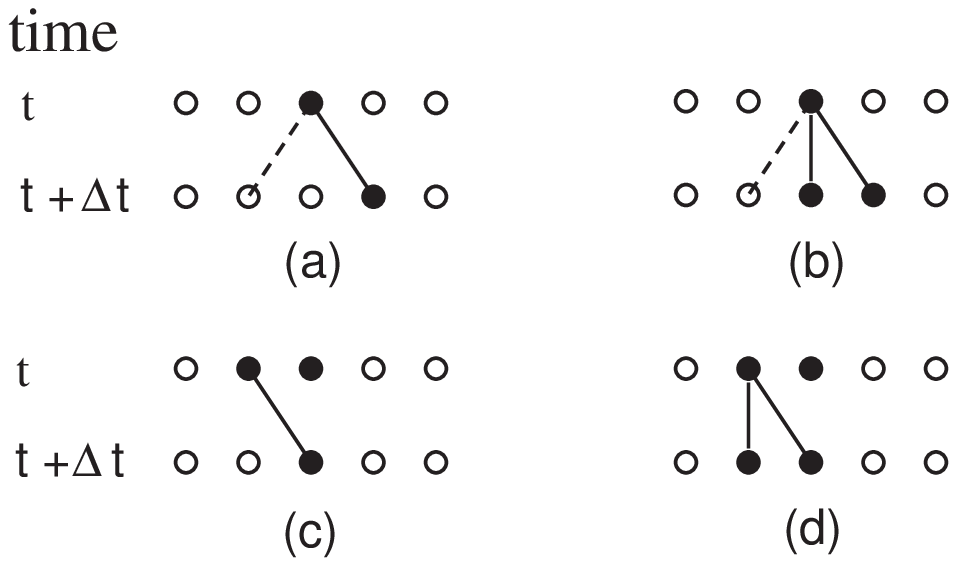}}
\noindent
{\small {\bf Figure~2}. Reaction rules: (a) diffusion; (b)
birth; and coalescence, (c) following diffusion, and (d) following a birth
event.  The dotted lines in (a) and (b) indicate alternative target sites.}
\end{figure} 

The IPDF method used for the exact analysis relies on the key concept of
$E_{n,m}(t)$---the probability that sites $n,n+1,\cdots,m$ are empty at time
$t$.  The probability that site $n$ is occupied is 
\begin{equation}
\label{disc-conc}
{\rm Prob}({\rm site\ }n{\rm\ is\ occupied})=1-E_{n,n}\;.
\end{equation}
The event that sites $n$ through $m$ are empty (prob. $E_{n,m}$) consists of 
two cases: site $m+1$ is also empty (prob. $E_{n,m+1}$), or it is occupied. 
Thus the probability that sites $n$ through $m$ are empty, but site
$m+1$ is occupied is $E_{n,m}-E_{n,m+1}$.  With this (and with a similar rule for
when the particle is to the left of the empty segment) one can write down
a rate equation for the evolution of the empty interval
probabilities~\cite{DBH,Coalescence,Doering,AAreviews}:
\begin{eqnarray}
\label{mastereq}
&&{\partial E_{n,m}\over\partial t}
   = {D\over a^2}(E_{n,m-1}-E_{n,m})\nonumber\\
   &&- {D\over a^2}(E_{n,m}-E_{n,m+1})\nonumber\\
   &&- {D\over a^2}(E_{n,m}-E_{n-1,m})\nonumber\\
   &&+ {D\over a^2}(E_{n+1,m}-E_{n,m})\nonumber\\
   &&- {v\over a}[(E_{n,m}-E_{n,m+1})+(E_{n,m}-E_{n-1,m})]\;.
\end{eqnarray}
For example, the first term on the r.h.s. represents the event that sites
$n,\dots,m-1$ are empty and a particle at site $m$ hops to $m+1$, thus
increasing $E_{n,m}$.  The second term represents the decrease in $E_{n,m}$
when a particle at $m+1$ hops into the empty interval $n,\dots,m$,
etc.
Eq.~(\ref{mastereq}) is valid for $m>n$.  The special case of
$m=n$ yields the boundary condition
\begin{equation}
\label{discBC}
E_{n,n-1}=1\;.
\end{equation}
The fact that the $\{E_{n,m}\}$ represent {\it probabilities\/} implies the
additional condition that $E_{n,m}\geq 0$.  Finally, if the system is not 
empty then $E_{n,m}\to 0$ as $n\to-\infty$ and $m\to\infty$.  

Rather than working with the discrete equations, it is simpler to pass to the
continuum limit.  We write
$x=na$ and $y=ma$, and replace $E_{n,m}(t)$ with $E(x,y,t)$.  Letting $a\to
0$, Eq.~(\ref{mastereq}) becomes
\begin{equation}
\label{eqE}
{\partial E\over\partial t}=D({\partial^2\over\partial x^2}
  +{\partial^2\over\partial y^2})E - v({\partial E\over\partial x}
  -{\partial E\over\partial y}) \;,
\end{equation}
with the boundary conditions, 
\begin{eqnarray}
\label{BC1}
E(x,x,t) &=& 1\;,\\
\label{BCpositive}
E(x,y,t) &\geq& 0\;,\\
\label{BC0}
\lim_{{x\to-\infty\atop y\to+\infty}}E(x,y,t) &=& 0\;.
\end{eqnarray}

The concentration of particles is obtained using Eqs.~(\ref{disc-conc}) and
(\ref{discBC}), and passing to the continuum limit:
\begin{equation}
\label{conc}
\rho(x,t)=-{\partial E(x,y,t)\over\partial y}|_{y=x}\;.
\end{equation} 
It can also be shown~\cite{DBH,trapdrift} that the conditional joint
probability for having particles at $x$ and $y$ but none in between, is
\begin{equation}
\label{IPDF}
P_2(x,y,t)=-{\partial^2E(x,y,t)\over\partial x\,\partial y}\;.
\end{equation}
Given a particle at $x$, the probability that the next nearest particle to
its {\em right\/} is at 
$y$, {\it i.e.}, the ``forward" IPDF, is
\begin{equation}
\label{IPDF.f}
p(x,y,t)=\rho(x,t)^{-1}P_2(x,y,t)\;.
\end{equation}  
Likewise, the ``backward"
IPD, the probability that the next nearest particle to the {\em left\/} of
a given particle at $y$ is at $x$, is 
\begin{equation}
\label{IPDF.b}
q(x,y,t)=\rho(x,t)^{-1}P_2(x,y,t)\;.
\end{equation} 

Eq.~(\ref{eqE}) admits two homogeneous stationary solutions: 
\begin{equation}
E(x,y)=1\;, 
\end{equation}and
\begin{equation}
\label{E_eq}
E_{\rm eq}(x,y)=e^{-{v\over D}(y-x)}\;.  
\end{equation}
The first solution implies $\rho=0$,
while the second solution describes active {\em equilibrium\/}, with
\begin{equation}
\rho_{\rm eq}=v/D\equiv\gamma\;.
\end{equation}
These states correspond, respectively, to
the unstable and stable phases of Fisher's theory (Eqs.~\ref{rho=0}
and~\ref{rho=k1/k2}).  It is important to notice that in the equilibrium
phase the particles are distributed independently from each other.
Indeed, suppose that the particles are so distributed, at a homogeneous
concentration $\gamma$. Then, the probability that there are no particles in
an infinitesimal interval of length
$\Delta\xi$, is $(1-\gamma\Delta\xi)$.  Since in equilibrium the particles are
uncorrelated, the probability that a finite interval of length $\xi=y-x$ is
empty, is $(1-\gamma\Delta\xi)^{\xi/\Delta\xi}$.  The equilibrium empty
interval probability is then recovered by taking the limit $\Delta\xi\to 0$.
Notice also that in the equilibrium state the IPDF (both forward or
backward) is 
\begin{equation}
\label{IPDF.eq}
p_{\rm eq}(x,y,t)=q_{\rm eq}(x,y,t)=\gamma e^{-\gamma(y-x)}\;.
\end{equation}
This can be derived from
Eqs.~(\ref{IPDF})--(\ref{IPDF.b}), and (\ref{E_eq}), as well as from the
properties of uncorrelated, randomly distributed particles.

Suppose now that a system is prepared in the following initial state:  at
$x<0$ the particles are distributed as in the equilibrium distribution,
Eq.~(\ref{E_eq}), and at $x>0$ there are no particles (the unstable steady
state).  The initial concentration profile is then
\begin{equation}
\label{ICdiffuse}
\rho(x,0)={v\over D}[1-H(x)]=\left\{ \begin{array}{ll}
v/D,  & \mbox{$x<0$}\\
0,     & \mbox{$x>0$} \end{array} \right.
\end{equation}
where $H(\cdot)$ is the Heavyside step function.
Doering et al.~\cite{DBH} have shown that in this case the concentration
profile evolves as
\begin{equation}
\label{rho.diffuse}
\rho(x,t)={v\over2D}{\rm erfc}\big({x-vt\over\sqrt{4Dt}}\big)\;,
\end{equation}
where ${\rm erfc}(\cdot)=1-{\rm erf}(\cdot)$ is the complementary error
function~\cite{AS}.  That is, the initial step-wave propagates at speed $v$,
while at the same time the wave front broadens as $w\sim\sqrt{Dt}$.

Doering et al.~\cite{DBH} have also considered the same initial condition as
above, but when a particle is known for sure to be at the front's edge at
$x=0$:
\begin{equation}
\label{ICsure}
\rho(x,0)={v\over D}[1-H(x)]+\delta(x)\;.
\end{equation}
In this case the wave front acquires an additional peak which propagates at
the same speed and broadens in the same fashion as before:
\begin{eqnarray}
\label{rho.sure}
\rho(x,t)=&&{v\over2D}{\rm erfc}\big({x-vt\over\sqrt{4Dt}}\big)\nonumber\\
&&+{1\over\sqrt{4\pi Dt}}\exp\big[-{(x-vt)^2\over4Dt}\big]\;.
\end{eqnarray}

We now show that the distribution of particles in the
coalescence system follows a simple and beautiful pattern:  The particles
in the stable phase---which are initially distributed as
in equilibrium (Eq.~\ref{E_eq})---remain distributed in the same fashion
throughout the lifetime of the wave.  The noisiness of the system is
manifested in the motion of the leading particle (the one at the wave's
edge):  rather than moving at constant speed $c=v$, it performs a biased
random walk with average drift velocity $v$.  Thus, the broadening of the
wave front is the result of averaging over the shifting positions of the
leading particle in different realizations of the process (Fig.~1). For
simplicity, we first show that such an interpretation is consistent with the
exact solution of ref.~\cite{DBH}, and we defer the complete proof until
later.

Let the position $z$ of the leading particle be given by the distribution
probability density $p(z,t)$.  Let the other particles---those to the left of
the leading particle---remain distributed as in equilibrium {\em at all
times\/}.  The probability that the interval $(x,y)$ is empty at time
$t$ depends upon the location of the interval endpoints, $x$ and $y$, with
respect to that of the leading particle, $z$:
\begin{equation}
E(x,y,t)=\left\{ \begin{array}{ll}
1, & \mbox{$z<x<y$}\;,\\
0, & \mbox{$x<z<y$}\;,\\
e^{-\gamma(y-x)}, & \mbox{$x<y<z$}\;. \end{array}
\right.
\end{equation}
Hence, taking into account the distribution of $z$,
\begin{eqnarray}
\label{Eedge}
E(x,y,t)&=&\int_{-\infty}^x p(z,t)\,dz 
     + e^{-\gamma(y-x)}\int_y^{\infty}p(z,t)\,dz\nonumber\\
&=& P(x,t)+e^{-\gamma(y-x)}[1-P(y,t)]\;,
\end{eqnarray}
where in the last equation we introduced the definition
\begin{equation}
\label{defP}
P(z,t)\equiv\int_{-\infty}^zp(z',t)\,dz'\;.
\end{equation}
In terms of $p$, the concentration of particles (Eq.~\ref{conc}) becomes 
\begin{equation}
\label{conc.P}
\rho(x,t)=p(x,t)+\gamma[1-P(x,t)]\;.
\end{equation}
Notice also that the forward IPDF is
\begin{equation}
p(x,y,t)=\gamma e^{-\gamma(y-x)}{p(y,t)+\gamma[1-P(y,t)]\over
                                p(x,t)+ \gamma[1-P(x,t)]}\;,
\end{equation}
while the backward IPDF remains exactly as in equilibrium (Eq.~\ref{IPDF.eq}),
consistent with the postulated distribution of particles.

It is straightforward to verify that the proposed form of $E(x,y,t)$,
Eq.~(\ref{Eedge}), satisfies the boundary conditions
(\ref{BC1})--(\ref{BC0}). Putting $E(x,y,t)$ in Eq.(\ref{eqE}) we see that
it is also satisfied, provided that 
\begin{equation}
\label {eqP}
{\partial\over\partial t}P(z,t)=D{\partial^2\over\partial z^2}P(z,t)
   -v{\partial\over\partial z}P(z,t)\;.
\end{equation}
{}From the definition of $P$, we have the boundary conditions:
\begin{eqnarray}
\label{BC.P1st}
\lim_{z\to-\infty}P(z,t)&=&0\;,\\
\lim_{z\to\infty}P(z,t)&=&1\;,\\
P(z,t)&\geq& 0\;.
\label{BC.Plast}
\end{eqnarray}
Finally, if a particle is known for sure to be
present initially at $x=0$ (Eq.~\ref{ICsure}), that particle is clearly the
leading particle, and $p(x,0)=\delta(x)$.  Thus,
\begin{equation}
\label{IC.P}
P(z,0)=H(z)\;. 
\end{equation}

The solution to Eq.~(\ref{eqP}) which satisfies the boundary conditions
~(\ref{BC.P1st})--(\ref{BC.Plast}) and the initial condition~(\ref{IC.P}), is
\begin{equation}
P(z,t)={1\over 2}+{1\over 2}{\rm erf}\big({z-vt\over\sqrt{4Dt}}\big)\;.
\end{equation}
Thus, using Eq.~(\ref{conc.P}), we recover the prediction of Doering et al.,
Eq.~(\ref{rho.sure}).  The probability density function for the position of
the leading particle is particularly simple:
\begin{equation}
p(x,t)=(4\pi Dt)^{-1/2}\exp[-(x-vt)^2/4Dt]\;,
\end{equation}
and identical to that of a Brownian particle characterized by the diffusion
coefficient $D$ and subject to a drift $v$.   A moment's reflection shows
why this is the case:  The particle at the edge of the wave can step to the
right or left with equal probabilities, at rate $D/a^2$.  This explains
diffusion.  The drift is a result of the birth mechanism.  If the leading
particle gives birth onto the site to its left, the edge does not
move.  However, if the particle gives birth onto the site to its 
right then the edge moves to the right (at rate $v/a$).  Because of this
left/right asymmetry the walk performed by the edge is biased. ---What is
surprising, is that the particles trailing the edge remain distributed, on
average, exactly as in their initial equilibrium distribution!

Consider now the initial condition~(\ref{ICdiffuse}), where the leading
particle is not necessarily at $z=0$.  Because the particles are
distributed as in equilibrium, the leading particle is at
$z<0$ with probability
$p(z,0)=\gamma e^{\gamma z}$ (and at
$z>0$ with probability $0$).  This case may be regarded as a superposition
of systems of the previous type, where the leading particle is initially
surely at $z$.  Indeed, the linear combination
\begin{eqnarray}
\rho(x,t)=\int_{-\infty}^0&&\gamma e^{\gamma z}
\big\{{v\over2D}{\rm erfc}\big({x-z-vt\over\sqrt{4Dt}}\big)\nonumber\\
&&+{1\over\sqrt{4\pi Dt}}\exp\big[-{(x-z-vt)^2\over4Dt}\big]\big\}dz\;,
\end{eqnarray}
reproduces the known result of Eq.~(\ref{rho.diffuse}).

Up to this point we have merely found a distribution of particles which
happens to explain the known results for $E$ and for the concentration of
particles $\rho$. It is conceivable, however, that other distributions might
accomplish the same feat.  A complete description of a particle system
requires knowledge of the full hierarchy of $n$-point density-density
correlation functions $\rho_n(x_1,x_2,\dots,x_n,t)$: the joint probability
to find particles at $x_1,x_2,\dots,x_n$ at time $t$.  The
particle concentration which was studied above corresponds to the special
(limited) case of $n=1$.  The $n$-point correlation functions can too be
analyzed exactly through the IPDF method \cite{Doering}.  We shall now
describe the procedure and employ it for the conclusion of our proof.

Let
$E_n(x_1,y_1,x_2,y_2,\dots,x_n,y_n,t)$ be the joint probability density that
the intervals $[x_i,y_i]$ ($i=1,2,\dots,n)$ are empty at time $t$.  The
intervals are non-overlapping, and ordered: $x_1<y_1<\cdots<x_n<y_n$. Then,
the $n$-point correlation function is given by
\begin{eqnarray}
\label{rho_n}
&&\rho_n(x_1,\dots,x_n,t)=\nonumber\\
&&(-1)^n{\partial^n\over\partial y_1\cdots\partial y_n}
E_n(x_1,y_1,\dots,x_n,y_n,t)|_{y_i=x_i}\;.
\end{eqnarray}
Doering~\cite{Doering} has shown that in our coalescence system the $E_n$
satisfy the partial differential equation:
\begin{eqnarray}
\label{dEn/dt}
{\partial\over\partial t}&& E_n(x_1,y_1,\dots,x_n,y_n,t)
   =\nonumber\\
&&D({\partial^2\over\partial x_1^2} +{\partial^2\over\partial y_1^2} +
   \cdots +{\partial^2\over\partial x_n^2} +
   {\partial^2\over\partial y_n^2})E_n\nonumber\\
&&-v[({\partial\over\partial x_1}-{\partial\over\partial y_1})+\cdots
   +({\partial\over\partial x_n}-{\partial\over\partial y_n})]E_n\;,
\end{eqnarray}
with the boundary conditions
\begin{eqnarray}
\label{bc:En.1}
&&\lim_{x_i\uparrow y_i{\rm\ or\ }y_i\downarrow x_i}
  E_n(x_1,y_1,\dots,x_n,y_n,t)=\nonumber\\
&&  E_{n-1}(x_1,y_1,\dots,\not{\!x_i},\not{\!y_i},\dots,x_n,y_n,t)\;,
\end{eqnarray}
and
\begin{eqnarray}
\label{bc:En.2}
&&\lim_{y_i\uparrow x_{i+1}{\rm\ or\ }x_{i+1}\downarrow y_i}
 E_n(x_1,y_1,\dots,x_n,y_n;t)=\nonumber\\
&& E_{n-1}(x_1,y_1,\dots,\not{\!y_i},\not{\!x_{i+1}},\dots,x_n,y_n;t)\;,
\end{eqnarray}
which should be obeyed for all $n>1$.
Here, we use the notation that crossed out arguments (e.g.
${\not{\!x_i}}$) have been removed.  Notice how the $E_n$ are tied together
in an hierarchical fashion through the boundary conditions~(\ref{bc:En.1})
and (\ref{bc:En.2}):  one must know $E_{n-1}$ in order to compute $E_n$.
At the root of the hierarchy, $E_1\equiv E$ is the simple empty interval
probability that was studied above.

If our interpretation of the particles distribution is correct, then,
following a reasoning similar to that which led to Eq.~(\ref{Eedge}), we
should have
\begin{eqnarray}
\label{En}
E_n&&(x_1,y_1,\dots,x_n,y_n,t)=P(x_1,t)\nonumber\\
&&+e^{-\gamma(y_1-x_1)}[P(x_2,t)-P(y_1,t)]\nonumber\\
&&+\cdots+e^{-\gamma\{(y_1-x_1)+\cdots+(y_i-x_i)\}}[P(x_{i+1},t)-P(y_i,t)]
                                           \nonumber\\
&&+\cdots+e^{-\gamma\{(y_1-x_1)+\cdots+(y_n-x_n)\}}[1-P(y_n,t)]\;.
\end{eqnarray}
It is easy to confirm that these functions fulfill the boundary 
conditions~(\ref{bc:En.1}) and (\ref{bc:En.2}).
Eq.~(\ref{dEn/dt}) is also satisfied, provided that $P$ satisfies the
same equation as before, Eq.~(\ref{eqP}).  Using Eqs.~(\ref{conc.P}),
(\ref{rho_n}), and (\ref{En}), we find the $n$-point correlation function:
\begin{equation}
\rho_n(x_1,\dots,x_n,t)=\gamma^{n-1}\rho(x_n,t)\;.
\end{equation}
Thus, the joint probability density for
finding particles at $x_1<x_2<\cdots<x_n$ depends only on the rightmost
particle at $x_n$---it alone is represented by the overall concentration
profile.  The particles at $x_1,\dots,x_{n-1}$ are distributed randomly and
independently of each other, with the equilibrium density $\gamma$.  This
is exactly what one would expect from the particle distribution that we
have championed throughout this work, and it completes the proof to our
claim.

In summary, we have shown that in one dimension the broadening of the wave
front of Fisher waves in the coalescence system is truly due to the spread
of locations of the leading particle in different realizations of the
process.  This peculiar situation, where the particles remain
distributed as in the initial equilibrium configuration---and where all the
changes are only reflected in the position of the leading particle---was
recently found also in diffusion-limited coalescence in one dimension
in the presence of a trap~\cite{trapdrift}.  In fact, the proof in the
present article may be carried over to that situation with only minor
changes.  

Fisher waves in the coalescence
system have been studied numerically also in higher
dimensions~\cite{waves}.  The goal there was to explore the effect of
fluctuations intrinsic to  particle systems as a function of
dimensionality.  The wave front was found to broaden as a power of time,
with an exponent which decreases with increasing $d$ and becomes zero (as in
Fisher's mean-field description) above the critical dimension of $d=3$.  The
width of the wave front in ref.~\cite{waves} was measured from the
concentration profile, as averaged over an ensemble of different
realizations.  It would be interesting to repeat these simulations, but this
time to subtract the fluctuations in the position of the wave front between
different runs, as suggested from the present findings.  This could result in
a decrease of the width exponent, and possibly also in a decrease of the
critical dimension for the validity of the mean-field Fisher picture.

I thank Paul Krapivsky for stimulating discussions.


\end{multicols}


\begin{thebibliography}{99}
\bibitem{remark} Our notation here differs from previous work: we take the
birth rate to be $v/a$ rather than $v/2a$, to achieve a more aesthetic form of
the final result.
\end{thebibliography}
\end{document}